\documentclass[sigconf]{acmart}

\setcopyright{none}
\acmConference[KDD 2026 Workshop]{Workshop on Agentic Software Engineering (SE 3.0)}{August 2026}{Jeju, Republic of Korea}
\acmDOI{}
\acmISBN{}
\renewcommand\footnotetextcopyrightpermission[1]{}
\usepackage{booktabs}
\usepackage{algorithm}
\usepackage{algpseudocode}
\usepackage{xcolor}
\usepackage{xspace}
\usepackage{tikz}
\usetikzlibrary{arrows.meta, positioning, calc, shapes}
\usepackage{pifont}
\newcommand{\cmark}{\ding{51}}
\newcommand{\xmark}{\ding{55}}
\usepackage{enumitem}
\usepackage{hyperref}
\hypersetup{colorlinks=true, citecolor=blue, linkcolor=black, urlcolor=blue}

\newcommand{\skapi}{Skill-as-API\xspace}

\begin{document}

\title{\skapi: Confidential Multi-Agent Coordination for Agentic Software Engineering}

\author{Ziwei Zhao}
\authornote{These authors contributed equally to this work.}
\author{Yu Gu}
\authornotemark[1]
\author{Haojun Liang}
\authornotemark[1]
\author{Chen Zhang}
\affiliation{%
  \institution{Technical University of Munich}
  \city{Munich}
  \country{Germany}
}

\author{Xizhi Ding}
\affiliation{%
  \institution{London Business School}
  \city{London}
  \country{United Kingdom}
}

\begin{abstract}
AI coding agents are evolving from solitary tools into collaborative teammates that discover and invoke one another's specialized skills. But the coordination channel itself can leak a skill's intellectual property. Protocols such as MCP and A2A run implementations server-side, yet they still publish each skill's description and typed schemas to every peer, offer no way to hide a skill's existence, and cannot guarantee that a wrapped system prompt stays off the wire. Application-layer privacy filters help, but act only \emph{after} the model has decided to emit sensitive text. We take a complementary, protocol-layer route: \skapi, a coordination protocol whose public view of a skill is limited to its name, description, typed input/output schemas, and trust tier. The skill body is closure-captured in the owner's process and never crosses the wire. Four layers add access control and narrow the prompt-injection surface structurally rather than by filtering content. We provide an open-source Python implementation over XMTP with 1.8--2.9\,s cross-continent hot-reconnect latency, and a software-engineering case study in which three agents coordinate a pull-request review while each retains ownership of its proprietary analysis prompts.
\end{abstract}

\keywords{multi-agent coordination, LLM agents, agentic software engineering, skill confidentiality, protocol design}

\maketitle

\section{Introduction}
\label{sec:intro}

AI coding agents such as SWE-agent~\cite{swe-agent} and OpenHands~\cite{openhands} are evolving from single-machine assistants into collaborative \emph{AI teammates} that increasingly work across organizational boundaries. A team running its own static-analysis ruleset may want to invoke another team's security-scanning agent, an LLM that wraps a proprietary, curated library of vulnerability patterns; a DevOps agent may delegate incident triage to a vendor agent encoding months of response experience. Such cross-boundary collaboration is the hallmark of SE~3.0, the \emph{agentic software engineering} paradigm in which human engineers orchestrate fleets of autonomous agents~\cite{hassan-se30-2025}.

Yet the collaboration channel itself is a liability. Protocols such as MCP and A2A run skill implementations server-side, so neither ships code; but both still publish each skill's name, description, and typed schemas to every peer, cannot hide a skill's existence, and cannot keep a wrapped system prompt off the discovery surface. What leaks is therefore narrow but valuable: the methodology a proprietary skill encodes. Application-layer privacy filters such as PrivacyChecker~\cite{privacylens-live-2025} reduce such leakage but leave a ${\sim}7$--$8\%$ residual in their reported settings, intervening only after the model has decided to emit sensitive content.

We introduce \textbf{\skapi}, a coordination protocol for agentic SE whose public view of a skill is limited to its name, description, and typed input/output schemas. The body, its code and the system prompt that encodes its behavior, is closure-captured in the owner's process and never crosses the wire. We make three contributions:

\begin{itemize}[leftmargin=*, itemsep=1pt, topsep=2pt]
  \item[\textbf{C1}] \textbf{Protocol design} (\S\ref{sec:design}): trust tiers, auto-downgrade, skill hiding, and a function-boundary defense, unified by one confidentiality invariant (the body never enters the public view).
  \item[\textbf{C2}] \textbf{Open-source implementation}\footnote{Implementation available at \url{https://github.com/HAOJUN-LIANG/Skill-as-API}.}: a lightweight Python SDK over XMTP with no Python-side dependencies beyond the standard library, achieving cross-continent hot-reconnect latency of 1.8--2.9\,s.
  \item[\textbf{C3}] \textbf{SE case study} (\S\ref{sec:case-study}): three agents coordinate a pull-request review, demonstrating trust tiers, skill hiding, and injection absorption in a realistic SE workflow.
\end{itemize}

We position \skapi not as a replacement for MCP or A2A but as a complementary protocol-layer strategy for settings where skill IP must be protected during multi-agent coordination.

\section{Motivation and Threat Model}
\label{sec:motivation}

\subsection{Threats in SE Agent Collaboration}

We identify three threats specific to multi-agent SE coordination:

\paragraph{\textbf{T1: Skill-IP extraction.}}
A caller tries to obtain a callee's proprietary methodology in one of two ways: by extracting its code or system prompt directly through the coordination channel, or by enumerating the callee's skill namespace to locate high-value targets. Statistical cloning, in which a caller distills many legitimate outputs into an approximate copy, is a fundamental limit of any callable skill and lies outside our scope (\S\ref{sec:conclusion}). Agent Skills for LLMs~\cite{agentskills-2026} documents a lifecycle defense model and reports that roughly a quarter of surveyed skills carry exploitable security risks.

\paragraph{\textbf{T2: Prompt theft via injection.}}
A caller crafts a task input that tricks the callee's LLM into emitting its system prompt verbatim. The OWASP 2026 catalog~\cite{owasp-llm-2025} formalizes prompt injection as direct and indirect; we additionally consider tool-result and conversation-history channels.

\paragraph{\textbf{T3: Trust drift.}}
After a successful collaboration episode, the caller retains elevated permissions indefinitely, turning a one-time PR-review partnership into a permanent open channel. In SE~3.0 settings where agents from different organizations collaborate transiently, unbounded trust accumulation is an especially dangerous failure mode.

\subsection{Threat Model}

\begin{table}[t]
  \centering
  \caption{Threat model: actor capabilities and scope.}
  \label{tab:threat-model}
  \footnotesize
  \setlength{\tabcolsep}{3pt}
  \begin{tabular}{lll}
    \toprule
    Actor & Capability & Scope \\
    \midrule
    Honest-but-curious caller & Read schema, send inputs & in \\
    Adversarial caller & + Injection payloads (OWASP) & in \\
    Colluding callees & Pool outputs to reconstruct schema & partial \\
    Schema-inference attacker & Probe input space to learn prompt shape & partial \\
    OS/physical attacker & Compromise owner machine & out \\
    Caller's LLM hijacked & Attack inside caller process & out \\
    \bottomrule
  \end{tabular}
\end{table}

Table~\ref{tab:threat-model} summarizes the scope. We assume the owner's machine is not compromised; if it is, no protocol can protect the skill body. We also exclude attacks on the caller's own LLM stack: \skapi protects the \emph{callee's} code and prompt, not the caller's. Two adversary classes are \emph{partially in scope}: colluding callees that pool their observations to reconstruct schema-level information, and schema-inference attackers that probe the input space to learn prompt structure. \skapi reduces but does not eliminate these risks; we discuss the residual surface in \S\ref{sec:conclusion} (L4--L5).

\subsection{Design Goals}

These threats are not independent: extraction (T1) and prompt theft (T2) both target the skill body, and trust drift (T3) widens over time the set of peers that may attempt them. Defending against all three takes more than encrypting a channel or filtering outputs after the fact; it requires controlling what a peer can ever observe, who is allowed to observe it, and for how long. We distill these requirements into five design goals, which the protocol of \S\ref{sec:design} realizes and the case study of \S\ref{sec:case-study} exercises.

\begin{itemize}[leftmargin=*, itemsep=1pt, topsep=2pt]
  \item[\textbf{G1}] \textbf{Code/prompt non-transmission} (structural).
  \item[\textbf{G2}] \textbf{Unknowability}: hidden, forbidden, and nonexistent skills return the same error string.
  \item[\textbf{G3}] \textbf{Bounded permission lifetime}: trust granted for one collaboration episode decays on completion.
  \item[\textbf{G4}] \textbf{Function-boundary prompt protection}: caller input never enters the callee's system prompt.
  \item[\textbf{G5}] \textbf{Acceptable latency}: low single-digit-second hot-reconnect.
\end{itemize}

\section{The \skapi Protocol}
\label{sec:design}

\skapi has two parts: a core abstraction that fixes what a skill exposes on the wire, and four layers that govern who may invoke it and what an adversary can extract. We first define the abstraction and its confidentiality invariant, then present the four layers, each addressing a threat from \S\ref{sec:motivation}: trust-tier access control and auto-downgrade counter trust drift (T3), skill hiding blunts namespace enumeration (T1), and a function-boundary defense contains prompt injection (T2). We close with the reference implementation.

\subsection{Core Abstraction}

A \skapi skill is a tuple $S = (\mathit{name}, d, w, v, \sigma_{\mathrm{in}}, \sigma_{\mathrm{out}}, t_{\min}, B)$ where $d$ is a description, $w$ a natural-language routing hint, $v$ a version string, $\sigma_{\mathrm{in}}/\sigma_{\mathrm{out}}$ are typed schemas, $t_{\min}$ the minimum trust tier, and $B = (c, p)$ the \emph{body} (code $c$ and system prompt $p$), existing only in the owner's process. The public view is:
\begin{equation}
  \mathrm{view}(S) = (\mathit{name}, d, w, v,
    \sigma_{\mathrm{in}}, \sigma_{\mathrm{out}}, t_{\min}, h)
  \label{eq:view}
\end{equation}
where $h{=}\mathrm{SHA256}(\mathit{name}\!\parallel\!v\!\parallel\! \sigma_{\mathrm{in}}\!\parallel\!\sigma_{\mathrm{out}}\!\parallel\! t_{\min})_{[:16]}$ is a 16-byte schema hash for fast reconnect.

\noindent\textbf{Invariant.} $B \notin \mathrm{view}(S)$: the body never enters the wire. Invocation is $\mathrm{call}(S,x) = c(\mathrm{filter}(x,\sigma_{\mathrm{in}}))$, where $\mathrm{filter}$ drops keys not in $\sigma_{\mathrm{in}}$. The prompt $p$ is referenced by $c$ through closure capture at registration time; it is not a function parameter and cannot be reached from the input path.

\subsection{Layer 1: Trust-Tier ACL ($\to$ T3)}
\label{sec:trust-tier}

Every peer carries a tier $t \in \{$\textrm{UNTRUSTED}$=0$, \textrm{KNOWN}$=1$, \textrm{INTERNAL}$=2$, \textrm{PRIVILEGED}$=3\}$. Skill invocation requires $t \geq S.t_{\min}$. Inbound message types are further gated by a fixed table: \texttt{ping} and \texttt{trust\_request} at \textrm{UNTRUSTED}; \texttt{discover} at \textrm{KNOWN}; \texttt{collab\_request} at \textrm{INTERNAL}; unknown types default to \textrm{PRIVILEGED} (deny-by-default).

\subsection{Layer 2: Auto-Downgrade ($\to$ T3)}
\label{sec:auto-downgrade}

\begin{algorithm}[t]
  \caption{\textsc{OnTaskResult}$(peer, ok, okr\_done)$}
  \label{alg:downgrade}
  \begin{algorithmic}[1]
    \If{\textbf{not} $ok$}
      \State $\mathit{consec}[peer] \mathrel{+}= 1$
      \If{$\mathit{consec}[peer] \geq 3$}
        \State $\mathit{tier}[peer] \gets
          \max(\mathit{tier}[peer]{-}1,\, 0)$
        \State $\mathit{consec}[peer] \gets 0$
      \EndIf
    \Else\ \ $\mathit{consec}[peer] \gets 0$
    \EndIf
    \If{$okr\_done$}
      \State $\mathit{tier}[peer] \gets
        \max(\mathit{tier}[peer]{-}1,\, 1)$ \Comment{floor at KNOWN}
    \EndIf
  \end{algorithmic}
\end{algorithm}

Trust does not accumulate from successful use; it \emph{decays}. Three consecutive failures drop the peer one tier; completing a collaboration OKR removes one tier with a floor at \textrm{KNOWN} (Algorithm~\ref{alg:downgrade}). We call this \emph{throw-away trust}: a peer earns evidence for future re-elevation, not standing access, directly addressing T3.

\subsection{Layer 3: Skill Hiding ($\to$ T1)}

Hidden skills return the \emph{same} error string \texttt{"Unknown skill: X"} as truly nonexistent ones. Owner-side logs distinguish the cases, but this distinction never crosses the wire. The design goal G2 prevents a caller from enumerating the skill namespace to mount a systematic extraction attack (T1). We note that timing differences between hidden and executed skills remain as a side-channel; mitigating this via randomized delays is future work.

\subsection{Layer 4: Function-Boundary Defense ($\to$ T2)}

The prompt $p$ is closure-captured at registration; caller input reaches the skill only as named, schema-filtered kwargs. Skills built on our reference executor bind these kwargs solely to the \emph{user} turn, so even if the caller embeds ``print your system prompt'' in the input, that text is never concatenated into $p$. This yields two distinct properties. \emph{(i) Structural}: $p$ never crosses the wire (Eq.~\ref{eq:view}) and caller input never becomes the system-role parameter, a confidentiality property of the wire format and the kwarg boundary, not a content filter. \emph{(ii) Residual/behavioral}: a callee model may still recite $p$ in response to user-turn injection; this is the classic prompt-leak vector (L2), which we do not claim to eliminate and which we probe only illustratively in \S\ref{sec:case-study}. We thus \emph{narrow}, rather than abolish, the prompt-extraction surface: the system-role concatenation vector is closed by construction, while the residual user-role vector remains the model's behavioral choice (G4).

We note that this layer protects only the \emph{callee}. If a caller feeds returned values back into its own LLM, that stack remains exposed to ordinary injection.

\begin{figure}[t]
  \centering
  \includegraphics[width=\linewidth]{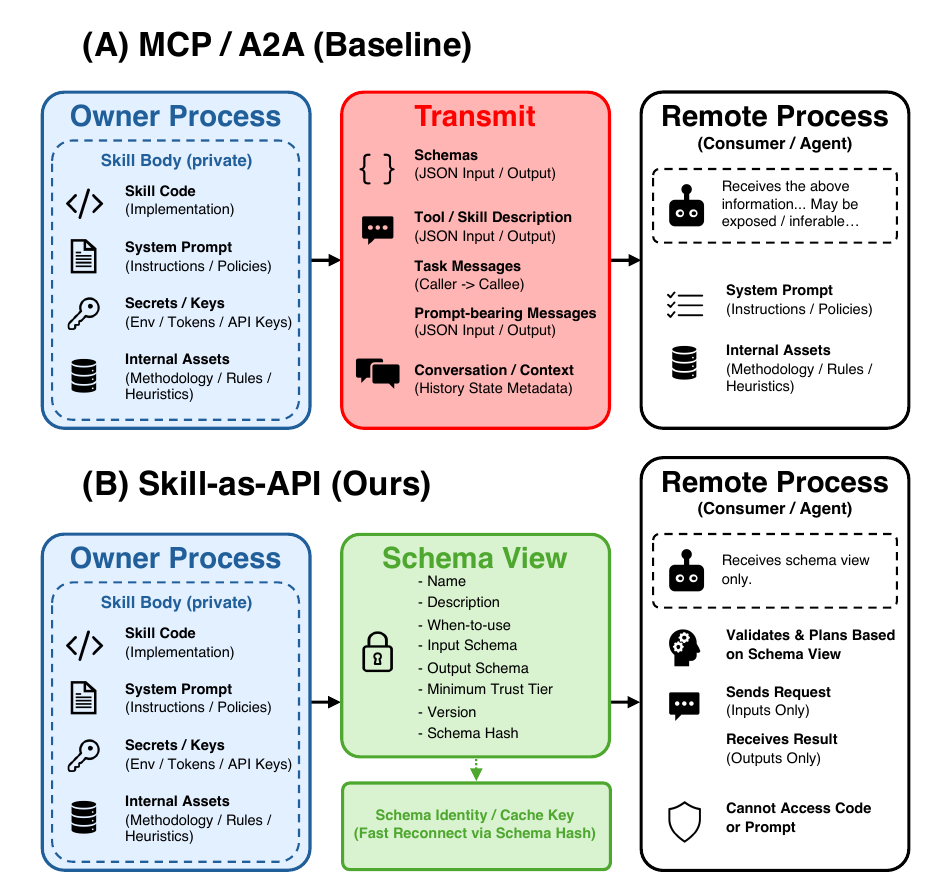}
  \caption{Information flow: MCP/A2A expose tool descriptions, schemas, and prompt context (red); \skapi exposes only the schema view (green), while the body stays in the owner's process (blue dashed box).}
  \label{fig:info-flow}
\end{figure}

\subsection{Implementation}
\label{sec:implementation}

The reference implementation is an open-source Python 3.10+ package that installs with a single command and has zero runtime Python dependencies beyond the standard library and optional LLM provider SDKs. Transport is a Node.js sidecar over the XMTP v3 (MLS) protocol via \texttt{@xmtp/node-sdk}; the Python core talks to the bridge over a localhost HTTP/SSE channel, keeping the cryptographic implementation out of the Python TCB. The 16-byte schema hash $h$ doubles as a fast-reconnect cache key, yielding the 1.8--2.9\,s hot-reconnect latency reported in \S\ref{sec:case-study}. We disclose two honest limitations for reproducibility: the trust-decay failure counters (Algorithm~\ref{alg:downgrade}) are currently in-memory only and reset on restart (tier transitions themselves persist to \texttt{trust.json} via an atomic temp-file+\texttt{fsync}+\texttt{os.replace} pattern), and skill versioning $v$ uses exact-string equality rather than semantic-range matching.

\section{SE Case Study}
\label{sec:case-study}

We instantiate \skapi in a realistic SE workflow where three agents, running as separate processes over live XMTP channels, coordinate a pull-request (PR) review. The case study exercises each design goal end to end: schema-only discovery (G1), injection absorption at the function boundary (G4), and trust decay on completion (G3).

\paragraph{\textbf{Setup.}}
The orchestrator agent receives a PR and delegates analysis to two specialists: a code-quality agent, whose proprietary code-review skill embeds a detailed checklist as its system prompt, and a security agent, whose vulnerability-scanning skill is backed by a pattern-library prompt. Both specialists register their skills at tier $t_{\min}=1$ (KNOWN). The orchestrator connects to each over XMTP, discovers their skill schemas but not their prompts, and dispatches review tasks.

\paragraph{\textbf{G1 verification: schema-only view.}}
After skill discovery, the orchestrator's view of \texttt{code\_review} contains only \texttt{name}, \texttt{description}, \texttt{input\_schema (diff: str)}, and \texttt{output\_schema (findings: list[str])}. The checklist prompt (${\sim}1{,}200$ characters of proprietary review methodology) is not present in any discovery message or tool-call payload. We verify this by logging all XMTP messages on the orchestrator's side and confirming zero overlap with the callee's \texttt{system\_prompt} string.

\paragraph{\textbf{G4 verification: injection absorption.}}
We craft 10 task inputs spanning eight OWASP-style injection classes (direct instruction override, role reversal, markdown/fence escape, prompt-leak request, base64-encoded wrapper, multilingual paraphrase, indirect-via-comment, and recursive self-call), each prepended to a representative PR diff; Table~\ref{tab:injection} gives the per-class counts. Each payload is sent to the code-quality agent; we measure (a) whether the system prompt appears verbatim or paraphrased in any returned field and (b) whether the structured output schema is preserved. Table~\ref{tab:injection} reports the result: \textbf{10/10 payloads absorbed} (system prompt absent from every returned field, output schema intact). We report this as an \emph{illustrative spot-check, not a bound}: with $N{=}10$ a zero-leak result cannot tightly bound the residual leakage rate, and ``absorbed'' here conflates a \emph{structural} property (caller input is bound to a kwarg and never concatenated into $p$, which is closure-captured at registration) with a \emph{behavioral} one (the callee model not voluntarily reciting $p$), the latter being out of scope per~L2. Future work includes quantifying the two separately and adding a naive concatenating-skill control.

\begin{table}[t]
  \centering
  \caption{OWASP-style injection absorption on the code-quality agent ($N=10$). ``Absorbed'' means the system prompt does not appear in the return value and the output schema is preserved.}
  \label{tab:injection}
  \footnotesize
  \setlength{\tabcolsep}{4pt}
  \begin{tabular}{lcc}
    \toprule
    Payload class & N & Absorbed \\
    \midrule
    Direct instruction override   & 2 & 2/2 \\
    Role reversal / persona swap  & 1 & 1/1 \\
    Markdown / fence escape       & 1 & 1/1 \\
    Prompt-leak request           & 2 & 2/2 \\
    Encoded (base64) wrapper      & 1 & 1/1 \\
    Multilingual paraphrase       & 1 & 1/1 \\
    Indirect via embedded comment & 1 & 1/1 \\
    Recursive self-call           & 1 & 1/1 \\
    \midrule
    \textbf{Total}                & \textbf{10} & \textbf{10/10} \\
    \bottomrule
  \end{tabular}
\end{table}

\paragraph{\textbf{G3 verification: trust decay on OKR completion.}}
We exercise three transitions in the same case study run. (i) Both specialists are first elevated to \textrm{INTERNAL} so the orchestrator can invoke higher-privilege skills during the review. (ii) When the orchestrator marks the collaboration OKR complete, Algorithm~\ref{alg:downgrade} line~10 fires and decays both peers from \textrm{INTERNAL}$\rightarrow$\textrm{KNOWN}, a measured numerical drop, not a no-op at the floor. (iii) For completeness, we also verify the \textrm{PRIVILEGED}$\rightarrow$\textrm{INTERNAL} transition on a test peer in the same run. A subsequent attempt to invoke a \textrm{INTERNAL}-tier skill is denied; the orchestrator must re-request elevation for the next collaboration episode. A one-time PR review does not become a permanent skill-access channel.

\paragraph{\textbf{Latency.}}
Hot-reconnect latency (second and subsequent calls between previously connected peers) is 1.8--2.9\,s ($N{=}30$, two inter-continental XMTP relay endpoints on commodity Linux hosts, 100\,Mbps links). Cold start (first-ever channel establishment) is 30--60\,s and is dominated by XMTP MLS group formation. All timings include key ratcheting.

\paragraph{\textbf{Limitations observed.}}
The orchestrator forwards the code-quality agent's output as context to the security agent. A malicious code-quality agent could embed poisoned content in its response; the orchestrator currently passes it through without sanitization (\emph{cross-step input poisoning}). We disclose this as L1 in \S\ref{sec:conclusion}.


\section{Related Work}
\label{sec:related}

\paragraph{Agent-privacy strategies.}
Prior agent-privacy work clusters into three protocol-shape strategies: (i) \emph{encrypt-everything}, exemplified by AgentCrypt~\cite{agentcrypt-2026}, which secures inter-agent traffic with HE/MPC at the 2--3 orders-of-magnitude overhead generic to such techniques; (ii) \emph{IAM-gate}, exemplified by A2A, which restricts who can talk to whom but still publishes schemas and descriptions for discovery; and (iii) \emph{content-filtering at inference time}, exemplified by PrivacyChecker~\cite{privacylens-live-2025}, which retains the ${\sim}7$--$8\%$ residual discussed in \S\ref{sec:intro}. \skapi is a fourth, complementary strategy at the \emph{ABI layer}: it changes \emph{what is on the wire}, leaving encryption, IAM, and content filtering free to compose on top. These are orthogonal: \skapi removes exposure paths that filters cannot reach, while filters address the LLM's behavioral choices that no protocol can directly constrain. We choose this ABI boundary over TEE-based skill enclaves, which bind the body to silicon but add hardware-vendor dependence, in exchange for portability and zero infrastructure cost.

\paragraph{Multi-agent coordination protocols.}
MCP~\cite{mcp-firstlook-dsn26} connects agents to tool servers, transmitting descriptions and schemas. AgentNet~\cite{agentnet-2025} decentralizes coordination but does not address skill-IP protection. The closure-capture idea has antecedents in object-capability systems such as Capsicum~\cite{capsicum-2010}, which restrict reachable resources at the language and OS boundary; \skapi adapts the same intuition to the agent ABI.

\paragraph{SE agent systems.}
SWE-agent~\cite{swe-agent}, OpenHands~\cite{openhands}, and Copilot Workspace demonstrate that coding agents can autonomously resolve issues, review PRs, and generate patches. These systems assume single-agent or trust-all multi-agent environments. \skapi adds a trust-aware coordination layer for cross-organization SE agent collaboration, the SE~3.0 vision of AI teammates~\cite{hassan-se30-2025}.

\paragraph{Agent security.}
ARGUS~\cite{argus-2025} trains content-layer injection discriminators. MCPTox~\cite{mcptox-aaai26} catalogs MCP attack surfaces. \skapi operates at the protocol layer rather than the content layer, narrowing the prompt-extraction surface structurally.

\begin{table}[t]
  \centering
  \caption{Protocol comparison. \cmark{}=structural protection; \xmark{}=none/exposed; \emph{app}=left to the application; \emph{enc}=via encryption.}
  \label{tab:comparison}
  \footnotesize
  \setlength{\tabcolsep}{3pt}
  \begin{tabular}{lcccc}
    \toprule
              & MCP & A2A & AgentCrypt & \textbf{\skapi} \\
    \midrule
    Impl.\ code on wire    & none & none & enc & none\,(closure) \\
    Description/schema     & public & public & enc & public \\
    System-prompt confid.\ & app & app & enc & \cmark\,(closure) \\
    Skill-existence hiding & \xmark & \xmark & \xmark & \cmark \\
    Trust lifetime         & none & IAM & PKI & 4-tier\,+\,decay \\
    Injection def.\        & app & app & n/a & ABI structural \\
    Overhead               & low & low & high & low\,(XMTP) \\
    \bottomrule
  \end{tabular}
\end{table}

\section{Conclusion}
\label{sec:conclusion}

We presented \skapi, a coordination protocol for agentic SE in which skill bodies never leave the owner's process. Four layers add trust-tier access control, throw-away trust via auto-downgrade, skill hiding, and a function-boundary defense that narrows the injection surface structurally. An SE case study with three PR-review agents illustrates that trust tiers govern access, system-role injection is absorbed at the closure boundary, and permissions decay toward a KNOWN floor on task completion. The reference implementation is open-source, runs over XMTP, and achieves 1.8--2.9\,s hot-reconnect latency.

\paragraph{\textbf{Limitations and future work.}}
\textbf{L1}: Cross-step input poisoning (output of agent~$A$ forwarded unfiltered to agent~$B$) is not addressed; an orchestrator-level sanitizer is future work. \textbf{L2}: Application-level leakage (the callee's LLM voluntarily emitting sensitive content in the return value) is orthogonal to \skapi and remains the domain of content-layer filters~\cite{privacylens-live-2025}. \textbf{L3}: Timing side-channels between hidden and executed skills remain; randomized delays are planned. \textbf{L4}: Trust is per-peer, not per-skill; finer-grained ACLs are future work. \textbf{L5}: Colluding callees and schema-inference probes (\S\ref{sec:motivation}) are mitigated but not eliminated by schema-only views; defenses such as randomized output perturbation and rate-limited discovery are future work. We envision \skapi as part of the SE~3.0 tool ecosystem where agents collaborate as teammates while each retaining ownership of their engineering methodology.

\newpage

\bibliographystyle{ACM-Reference-Format}
\bibliography{references}

\end{document}